\documentclass[useAMS,usenatbib]{mn2e}
\usepackage{times}
\def\spose#1{\hbox to 0pt{#1\hss}}
\newcommand\lsim{\mathrel{\spose{\lower 3pt\hbox{$\mathchar"218$}}
     \raise 2.0pt\hbox{$\mathchar"13C$}}}
\newcommand\gsim{\mathrel{\spose{\lower 3pt\hbox{$\mathchar"218$}}
     \raise 2.0pt\hbox{$\mathchar"13E$}}}

\begin{document}

\title[Cooling function of HD molecule]{The cooling function of HD molecule
revisited}
\author[Lipovka et al.]{Anton Lipovka$^{1}$\thanks{%
E-mail: aal@cajeme.cifus.uson.mx }, Ramona N\'u\~nez-L\'opez$^{1,2}$, and Vladimir 
Avila-Reese$^{2}$ \\
$^{1}$Centro de Investigaci\'on en F\'{\i}sica, UNISON, Rosales y Blvd. 
Transversal, Col. Centro, Edif. 3-I, Hermosillo, Sonora, M\'exico, 83000 \\
$^{2}$Instituto de Astronom\'{\i}a, U.N.A.M., A.P. 70-264, 04510, M\'exico,
D.F., M\'exico}

\maketitle

\begin{abstract}
We report new calculations of the cooling rate of primordial gas by the HD
molecule, taking into account its ro-vibrational structure. The HD cooling
function is calculated including radiative and collisional transitions for $%
J\le8$ rotational levels, and for the vibrational levels $v=0,1,2$ and 3.
The ro-vibrational level population is calculated from the balance equation
assuming steady state. The cooling function is evaluated in the ranges of
the kinetic temperatures, $T_k$, from $10^2$ to $2\times 10^{4}$ K and the number
densities, $n_H$, from 1 to $10^{8}$ cm$^{-3}$. We find that the inclusion
of collisional ro-vibrational transitions increases significantly the HD
cooling efficiency, in particular for high densities and temperatures. For $%
n_H\gsim 10^5$ and $T_k\sim 10^{4}$ K the cooling function becomes more than
an order of magnitude higher than previously reported. We give also the
HD cooling rate in the presence of the cosmic microwave radiation field for
radiation temperatures of 30, 85 and 276 K (redshifts of 10, 30 and 100). The
tabulated cooling functions are available at
http://www.cifus.uson.mx/Personal\_Pages/anton/DATA/HD\_cooling/HD\_cool.html. 
We discuss the relevance to explore the effects of including our results into
models and simulations of galaxy formation, especially in the regime when gas
cools down from temperatures above $\sim 3000$K.
\end{abstract}

\pagerange{\pageref{firstpage}--\pageref{lastpage}} \pubyear{2002}

\label{firstpage}


\section{Introduction}

The cooling and thermal balance of the primordial, zero-metallicity gas by
molecular hydrogen (H$_2$) and deuterated hydrogen (HD) are key ingredients
in the formation process of the first baryonic objects in the universe, such as
dwarf-sized galaxies and Pop-III stars (see for recent reviews on this topic
Barkana \& Loeb 2001; Ciardi \& Ferrara 2004; Bromm \& Larson 2004). It is
well known that H$_2$ and HD molecules form in the universe after
recombination (e.g., Lepp \& Shull 1984; Puy et al. 1993; Palla, Galli, \&
Silk 1995; Galli \& Palla 1998; Stancil, Lepp, \& Dalgarno 1998). The
formation of H$_2$ and HD molecules in primordial gas is also possible in a
postshock flow (Mac Low \& Shull 1986; Shapiro \& Kang 1987; Uehara \&
Inutsuka 2000), where the gas density and temperature, as well as the
molecular fractions, are different from those of the expanding homogeneous gas.

The H$_2$ molecule is considered typically as the main coolant in the
primordial medium (e.g., Tegmark et al. 1997; Abel, Bryan \& Norman 2000;
Bromm, Coppi \& Larson 2002). Thus, rather detailed calculations were
carried out for it (see for the most recent results and references
therein Le Bourlot, Pineau des For\^ets \& Flower 1999; Flower et al. 2000;
Shaw et al. 2005). Nevertheless, in the case of dense primordial gas, the
role of the HD molecule in the thermal and cooling process may become
comparable to or more important than the one of H$_2$. Therefore, detailed
calculations of the HD cooling function (the rate of cooling per HD
molecule, hereafter CF) are needed.

The HD molecule is an efficient coolant in the primordial medium because it
has a permanent electric dipole moment that allows high probabilities
for the radiative rotational transitions from the rotational levels pumped
by collisions with H and He. Besides the HD molecule was efficiently
produced in the reaction of chemical fractionation $D^{+}+H_{2}\rightarrow \
H^{+}+HD$, which leads to a strong enhancement of the initial abundance of
HD, up to roughly $[HD]/[H_{2}]=10^{-2}-10^{-3}$ (e.g. Puy et al. 1993;
Galli \& Palla 1998; Flower 2000). The HD molecule can cool the primordial
gas to temperatures below $\sim 300$ K because it has allowed dipole
rotational transitions characterized by energies two times smaller than
those of the quadrupole transitions of H$_{2}$. In that regard higher
temperatures ($\gsim 3000$K), cooling by HD can be again as important as that
by H$_{2}$. This question will be discussed in the present paper.

The HD CF was calculated for the first time by Dalgarno \& McCray (1972) for low
temperatures and Boltzmann distributed rotational level populations.
Varshalovich \& Khersonskii (1976) improved the calculations by taking into
account the departure from local thermodynamic equilibrium. More recently,
Flower et al. (2000, and see also more references therein) have calculated 
the HD CF for low and intermedium kinetic temperatures and a large range of
densities. These authors considered the rotational transitions within the 
vibrational ground state $(v=0)$ of HD. Note that in the case of higher 
kinetic temperatures ($T_{k}\gsim 2\times 10^{3}K$), the populations of the 
vibrational levels become significant, and cooling by ro-vibrational transitions 
of HD may be important due to (i) the relatively high energy of vibrational 
quanta, and (ii) the relatively high probabilities of the ro-vibrational 
radiative transitions. Therefore, the inclusion of the the vibrational 
structure of the HD molecule in the calculation of its CF might be important. 

The aim of this paper is to recalculate the HD CF for a wide range of
kinetic temperatures and densities by taking into account ro-vibrational
transitions within ground $(v=0)$ and the first three excited $(v=1,2,3)$
vibrational levels. Further exploration of the influence of the HD CF
calculated here on the thermal evolution of the primordial medium will be
presented elsewhere. In \S 2 we give details of our calculations. The
revised HD CF is given and compared with that from Flower et al. (2000) in 
\S 3. Finally, in \S 4 we discuss our results and speculate about possible
implications.

\section{The Model}

It is well known that in the low density limit ($n_{H}<10^{3}$cm$^{-3}$), the
relative populations of the levels of HD molecule depart significantly from
the Boltzmann distribution. Therefore, to calculate the HD CF in the general
case, we need the correct values of the ro-vibrational level populations outside
the thermodynamical equilibrium. This can be calculated by using the
equations of detailed population level balance: 
\begin{eqnarray}
n_{vJ}\sum_{v^{\prime }J^{\prime }}(W_{vJ\rightarrow v^{\prime }J^{\prime
}}^{R}+W_{vJ\rightarrow v^{\prime }J^{\prime }}^{C}) &=& \cr %
\sum_{v^{\prime}J^{\prime }}n_{v^{\prime }J^{\prime }}(W_{v^{\prime
}J^{\prime } \rightarrow vJ}^{R}+W_{v^{\prime }J^{\prime }\rightarrow
vJ}^{C}); \cr \sum_{vJ}n_{vJ} &=&1,  \label{1}
\end{eqnarray}%
where $n_{vJ}$\ is the population of the ro-vibrational level $vJ$, $%
W_{vJ\rightarrow v^{\prime }J^{\prime }}^{R}$\ and $W_{vJ\rightarrow
v^{\prime }J^{\prime }}^{C}$\ are the probabilities of the radiative and
collisional transitions, respectively. We assume a steady state for the
population levels. This assumption has been used and discussed to be 
valid in our context by many other authors (e.g., Flower et al. 2000; Flower
\& Pineau des For\^ets 2001; Le Petit, Roueff \& Le Bourlot 2002).  

The population of the ro-vibrational levels of the $HD$ molecule was
calculated in the wide range of number densities $n_{H}$ from $1$ to $%
10^{8}\ $cm$^{-3}$, and for the kinetic temperature $T_{k}$ from $10^{2}$ to $%
2\times 10^{4}$ K. The ro-vibrational radiative transition probabilities for
the HD molecule were taken from Abgrall et al. (1982). Regarding the
collisional transition probabilities, it was shown in Flower et al. (2000)
that the HD CF is insensitive to the H/H$_{2}$ density ratio. Therefore, one
may take into account the excitation and de-excitation of HD only by
collisions with the H atoms. The probabilities of the collisionally excited 
pure rotational transitions of HD (up to the third vibratioanally excited 
state, $0\leq v \leq 4$,
and up to the 8-th rotational level $0\leq J\leq 8$) were taken from Roueff \&
Flower (1999) and Roueff \& Zeippen (1999). We consider only the dipolar 
transitions ($\Delta J=\pm 1$). 
Owing to the permanent dipolar moment of the HD molecule, these
transitions dominate collisional population transfer (Flower \& Roueff 1999). 

Flower \& Roueff (1999) computed probability coefficients for several 
ro-vibrational transitions in HD induced by collisions with atomic and molecular 
hydrogen in a range of temperatures from 100 to 2000 K. In that paper, 
some values of the transition probabilities for the vibrational relaxation 
$v = 1 \rightarrow 0$ are presented and compared with the corresponding 
coefficients for H$_{2}$. The rate coefficients for vibrational relaxation 
of HD and H$_2$ in collisions with H are similar in magnitude for a given 
temperature. Besides, the probabilities for both molecules change approximately
in the same way with temperature. On the basis of these results and in the absence 
of extensive calculations for the HD+H collisional ro-vibrational transition 
probabilities in the range of temperatures we want to explore, we will use 
the corresponding probabilities of the H$_{2}$+H (Tin\'e, Lepp \& Dalgarno 1998) 
given in electronic form on the website
http://www.physics.unlv.edu/astrophysics/h2h2rates/index.html   
(Lepp, Tin\'e \& Dalgarno 1997, in preparation).

The total CF per unit of volume is defined, for example for the HD molecule,
as: 
\begin{equation}
\Lambda _{HD}=n_{HD}W_{HD},  \label{2}
\end{equation}%
where $n_{HD}$ is the HD number density, and $W_{HD}$ is the HD CF in
unities of erg s$^{-1}$ per HD molecule: 
\begin{equation}
W_{HD}=\sum_{vJv^{\prime }J^{\prime }}(n_{vJ}W_{vJ\rightarrow v^{\prime
}J^{\prime }}^{R\downarrow }-n_{v^{\prime }J^{\prime }}W_{v^{\prime
}J^{\prime }\rightarrow vJ}^{R\uparrow })h\nu _{vJ\rightarrow v^{\prime
}J^{\prime }}.  \label{3}
\end{equation}%
Here, $W_{vJ\rightarrow v^{\prime }J^{\prime }}^{R\downarrow
}=A^{R\downarrow }+B^{R\downarrow }\mathbf{u}$\ are the probabilities of the
radiative transitions with the emission of the photon $h\nu _{vJ\rightarrow
v^{\prime }J^{\prime }}$, and $W_{v^{\prime }J^{\prime }\rightarrow
vJ}^{R\uparrow }=B^{R\uparrow }\mathbf{u}$\ are the probabilities of
radiative transition with the absorption of the corresponding field photon.
The symbols $A$ and $B$ are for the corresponding Einstein coefficients and $%
\mathbf{u}$ is the radiation field. The populations of the ro-vibrational
levels $n_{vJ}$ are those calculated from the balance equation (1). This
definition of $W_{HD}$ is more general than the one used in Flower et al.
(2000), because it takes into account the potential effects of a radiation
field in the cooling process. In the case of $T_{r}<<T_{k}$, eq. (3) tends
to the case considered in Flower et al. (2000). At high redshifts, the
Cosmic Microwave Background Radiation (CMBR) temperature, $T_{r}=T_{\rm CMBR}$,
may become comparable to the kinetic temperature of the gas $T_{k}$; this
will significantly affect the HD CF at that $T_{k}$ (see below for more
details).

\section{The cooling function of HD molecule}

In Fig. 1 we present the variation of $W_{HD}$ with $T_{k}$ and $n_{H}$. The
dotted-line curves correspond to the HD CF, neglecting the collisional
ro-vibrational transitions as in Flower et al. (2000). Our results agree
very well with those of these authors. Note that we did not consider
collisions of HD with He and H$_{2}$ because, as it was mentioned above,
Flower et al. (2000) showed that $W_{HD}$ is essentially a function of $T_{k}$
and $n_{H}$ only. The solid-line curves in Fig. 1 are the CFs including
ro-vibrational collisional transitions for a 4-level vibrational structure
of the HD molecule ($v=0,1,2,3$). One sees that the latter CFs depart from
the former, as $T_{k}$ and $n_{H}$ are larger. For $n_{H}\gsim10^{6}$cm$^{-3}$
and $T_{k}\approx 10^{4}$K, the calculated value of $W_{HD}$ including the
vibrational transitions is more than a factor of fifty larger than the $%
W_{HD}$ calculated neglecting collisional ro-vibrational transitions. Note
that in Fig. 1, $W_{HD}$ has been calculated for $T_{r}=2.73$ K (the CMBR
temperature at present), i.e., well within the limit $T_{r}<<T_{k}$,
when the radiation field does not participate in the thermal balance of the
gas and eq. (3) is reduced to the equation for $W_{HD}$ used in Flower et al.
(2000). This way were able to compare our results with those of these
authors.

The exact HD CF data presented here are available at 
http://www.cifus.uson.mx/Personal\_Pages/anton/ DATA/HD\_cooling/HD\_cool.html. 
To facilitate the use of the HD CF in computational programs
we give also a polynomial approximation for the HD CF. The approximation is
very accurate in the ranges of $n_{H}$ and  $T_{k}$ studied here, from 
$1$ to $10^{8}$ cm$^{-3}$, and from $10^{2}$ to $2\times 10^{4}$ K,
respectively. The approximation is writen in the form: 
\begin{equation}
Log(W_{HD})=\sum_{l,m=0}D_{lm}T_{k}^{l}n_{HD}^{m},  \label{4}
\end{equation}%
where the coefficients $D_{lm}$ are tabulated in Table 1.

\begin{table}
\caption{Polynomial coefficients $D_{lm}$}%
\begin{tabular}{@{}lccccc}
\hline
& $m=0$ & $m=1$ & $m=2$ & $m=3$ & $m=4$ \\ 
 \hline
$l=0$ & -42.57688 & 0.92433 & 0.54962 & -0.07676 & 0.00275 \\ 
$l=1$ &  21.93385 & 0.77952 & -1.06447 & 0.11864 & -0.00366 \\ 
$l=2$ & -10.19097 & -0.54263 & 0.62343 & -0.07366 & 0.002514 \\ 
$l=3$ &  2.19906 & 0.11711 & -0.13768 & 0.01759 & -0.000666317 \\ 
$l=4$ & -0.17334 & -0.00835 & 0.0106 & -0.001482 & 0.000061926\\
  \hline
\end{tabular}%
\end{table}

From Fig. 1 one clearly sees  how $W_{HD}$ increases at high densities after
including the collisional ro-vibrational level transitions. In the case of
low densities, the contribution of these transitions to the cooling of HD is
less important. Nevertheless, at high temperatures $W_{HD}$ is still larger 
than in the case when these transitions are omitted (pure rotational
transitions shown by dashed line at Fig. 1). Therefore, the analytical
approximation commonly used for the low density limit (see e.g., Galli \&
Palla 2002) should be slightly modified. The approximation given in Galli \&
Palla (2002) is based on the calculations for only two low rotational
transitions $(J^{\prime }J)=(10)$ and $(J^{\prime }J)=(21)$, because the
collisional probabilities of these transitions are rather large as compared
with those of higher rotational levels. However, it should be stressed that
when the vibrational structure is taken into account, then there are other
ro-vibrational transitions with comparable probabilities of collisional
excitation. Based on our results, we suggest the following modified
approximation for the low-density limit HD CF: 
\begin{eqnarray}
Log(W_{HD})=-42.45906+21.90083T_{k}^{{}}-10.1954T_{k}^{2}+ \cr%
2.19788T_{k}^{3}-0.17286T_{k}^{4}.
\end{eqnarray}
This aproximation can be applied for the gas
density $n_{H}$ up to $10^{3}-10^{4}$ cm$^{-3}$.

\begin{figure}
\vspace{11cm} \includegraphics{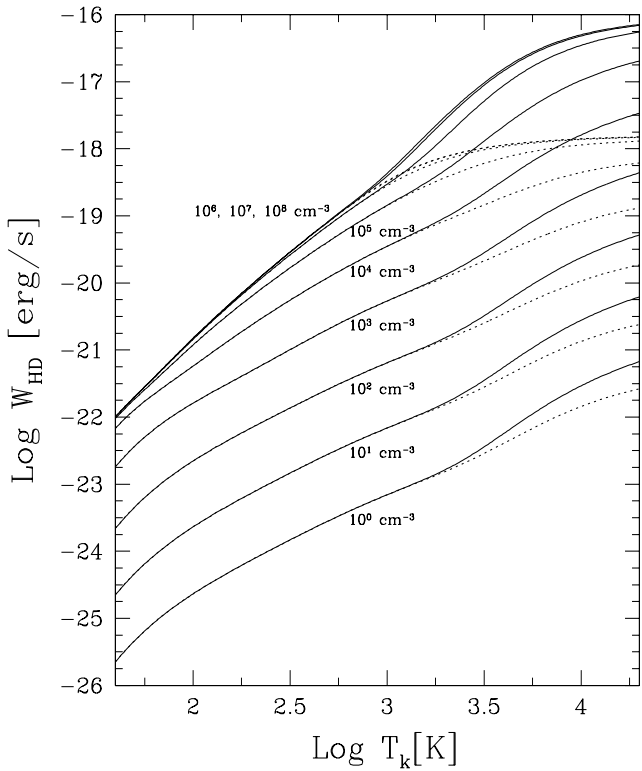}
\caption{Cooling function of the HD molecule, calculated from $n_{H}=1$ to $%
10^{8}$ cm$^{-3}$ in powers of ten (from bottom to top). Radiation
temperature $T_{r}$ was set equal to 2.73 K ($T_{r}<<T_{k}$). Solid curves
correspond to the cases when collisional ro-vibrational transitions ($%
v=0,1,2,3$, and $J\leq 8$) were taken into account, while dotted curves are
for the case when these transitions were neglected in the calculations. The
latter curves can be compared with those of Fig. 3 in Flower et al. (2000).
Note that the curves in both cases saturate for $n_{H}\gsim10^{5}-10^{6}$.}
\end{figure}

The effects of the radiation field $\mathbf{u}$ in $W_{HD}$ become important 
when $T_{r}\approx T_{k}$ (Flower 2000; Flower \& Pineau des For\^ets 2001). 
In this case, the radiative absorption by the HD molecule
dominates and the second term in eq. (3), as well as the $B^{R\downarrow }%
\mathbf{u}$ term in the probability $W_{vJ\rightarrow v^{\prime }J^{\prime
}}^{R\downarrow }$, should be considered. The second term in eq. (2) can be
interpreted as a heating function. At the redshfits at which the first baryon
objects are expected to form ($z\approx 10-100$), the CMB temperature can be
similar to the typical kinetic temperatures of the small
primordial gas clouds. In Fig. 2 we present the complete HD CFs (with the
collisional ro-vibrational transitions included) considering a radiative
field with $T_{r}=T_{\rm CMBR}(0)(1+z)$ for redshifts $z=10$, $30$ and $100$
(dotted, dashed and long-dashed lines, respectively) and for two densities, $%
n_{H}=10$ and $10^{8}$ cm$^{-3}$. One sees that when $T_{r}\approx T_{k}$, 
$W_{HD}$ falls dramatically. The second term in eq. (3) actually comes to 
dominate and the HD molecular lines act as heating sources. This will 
inhibit the collapse of a gas cloud. Therefore the minimum temperature of 
the primordial collapsing clouds is limited by the CMBR temperature (or by the 
heating due to other radiation fields) and not by HD (and H$_{2}$) cooling.

We note that the steady state assumption for the HD population levels (see 
\S 2) remains even at the high redshifts considered here. The typical 
time scales for population level changes are determined mainly by the inverse 
of the transition Einstein coefficients. For the HD molecule transitions
these coefficientes imply typical time scales much less than the time scales 
related to variations in $T_{\rm CMBR}$ and/or to the gas clouds 
collapse in the epochs considered here ($z\lsim 100$).

\begin{figure}
\vspace{10cm} \includegraphics{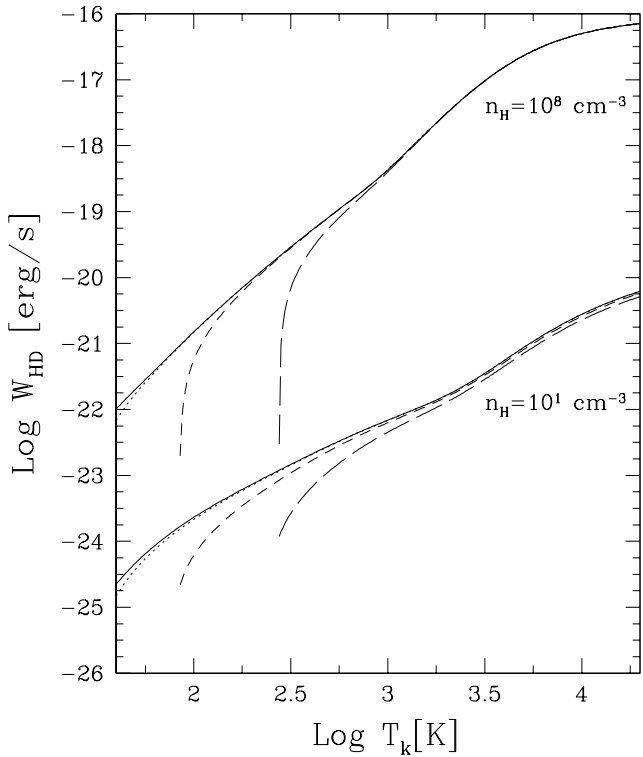}
\caption{Cooling function of the HD molecule (including collisional
ro-vibrational transitions) calculated for different values of the radiation
temperature, $T_r=2.73(1+z) K$, $z = 0$, $10$, $30$ and $100$ (solid, dotted,
dashed, and long-dashed lines, respectively), and for $n_H=10$ and $10^{8}$ 
cm$^{-3}$.}
\end{figure}

\section{Discussion}

We have calculated the HD CF, $W_{HD}$, for a wide range of kinetic
temperatures and gas densities by taking into account the ro-vibrational
structure of the molecule for both the radiative and collisional
transitions. We have found that, when including the collisional
ro-vibrational transitions, the cooling efficiency of HD molecule is higher
than previously reported. The main increasing of $W_{HD}$ as compared to
previous works is at the high-temperature side and for high densities (see
Fig. 1); for $n_H>10^5$ cm$^{-3}$, the differences in $W_{HD}$ reach an
order of magnitude and more at $T_k\sim 10^4$. Our results in support of
recent claims (Flower 2000; Uehara \& Inutsuka 2000; Flower \& Pineau
des For\^ets 2001; Nakamura \& Umemura 2002; Galli \& Palla 2002), suggest 
that the HD molecule may play an important role in the thermal balance and 
cooling of primordial gas at high densities.

In Fig. 3 we attempt to compare the relative contributions to the CF of the 
gas by the H$_2$ and HD molecules. Because the ratio of 
abundances of HD to H$_2$ is significantly smaller than 1, the $W_{HD}$ curves 
should be reduced by a factor equal to this ratio in order to compare the 
contributions to the cooling of gas by both molecules. As it was mentioned 
above, this ratio for the primordial gas, after chemical
fractionation, is approximately $10^{-2}-10^{-3}$. In Fig. 3, we reduce
$W_{HD}$ by a factor of $[HD]/[H_{2}]=10^{-2.4}$ according to Galli \& 
Palla (2002) for $z=10$. As one can see in Fig. 3, the role of
HD in the cooling of low-density gas is still unimportant with respect
to the one of  H$_2$, while for high densities, HD becomes as efficient 
a coolant as H$_2$, being even more efficient at low temperatures.
The H$_2$ CF has been calculated according to Flower et al. (2000).

\begin{figure}
\vspace{7cm} \includegraphics{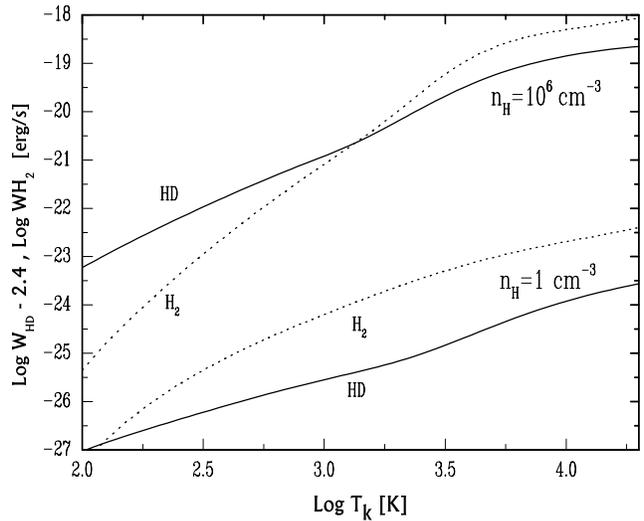}
\caption{Cooling functions of the HD (solid lines) and H$_2$
(dashed lines) molecules for two gas densities, $n_H = 1$ and $10^6$ cm$^{-3}$,
as indicated in the panel. To be compared, the HD CF was reduced by the
factor $[HD]/[H_{2}]=10^{-2.4}$. The H$_2$ CF is from Flower et al. (2000).}
\end{figure}

The contribution to the gas cooling of HD found here for high gas densities 
and temperatures larger than $\sim 3000$ K is close to the corresponding 
contribution of H$_2$. This will surely affect calculations of galaxy formation 
in the regime when dense gas cools down from temperatures above $\sim 3000$ K. 
This regime may arise in photoionization-heated gas or in shock-heated dense
material in dwarf galaxy-size dark matter halos with masses $\gsim 10^8$ M$%
_\odot$ (Bromm \& Larson 2004). The presence of an efficient coolant, 
besides that H$_2$,  able to operate at $T_k\sim 10000$ K 
and at high densities (as is the case of HD) is
probably also relevant to understand the formation of the first globular
clusters. Bromm \& Clarke (2002) have shown that, under the action of an
efficient coolant at $T_k\lsim 10000$ K, stellar clusters with masses $\sim
10^5-10^7$ may form into small dark matter subhalos that later on are
tidally destroyed during the violent relaxation of dwarf galaxy-sized halos
at $z\gsim 10$ (see also Weil \& Pudritz 2001).

The cooling efficiency of H$_{2}$ or HD molecules increases with density
until the population levels reachs the local thermodynamical equilibrium at
a critical density $n_{c}$, in such a way that at $n_{c}$ the probability of
collisional de-exitation becomes equal to the spontaneous radiative
probability. Beyond this density, the cooling is saturated. For molecular
hydrogen, the critical density is rather low, $n_{c,H_{2}}\sim 10^{3}-10^{4}$
cm$^{-3}$, due to the relatively small values of the H$_{2}$ Einstein
coefficients. In the case of the HD molecule, the Einstein coefficients are
larger as compared to those of H$_{2}$ by $2-3$ orders of magnitude. One may
estimate the ratio of critical densities $n_{c,HD}/n_{c,H_{2}}$ as $\approx
A_{v^{\prime }J^{\prime }vj}(HD)/A_{v^{\prime }J^{\prime }vj}(H_{2})$, which
is $\approx 10^{2}$ for radiative vibrational transitions. The value of $%
n_{c,HD}$ is indeed around two orders of magnitude higher than $n_{c,H_{2}}$%
. From Fig. 1 (see also Flower et al. 2000), one sees that $n_{c,HD}\approx
10^{5}-10^{6}$ cm$^{-3}$. To conclude, we remark the importance of
taking into consideration the HD CF calculated here in the simulations and 
models of the first baryonic objects in the universe. In particular, 
those processes related to dense gas that cool down from temperatures above 
$\sim 3000$ K will be
affected by our results.

\section*{ACKNOWLEDGMENTS}

We are thankful to the anonymous referee for his/her remarks
and comments that helped to improve the paper. 
This work was supported by grant PROMEP /103.5/03/1147 to
A.~L. In addition, R.~N-L. gratefully acknowledges a
PhD fellowship by CONACyT. We are grateful to J. Benda for 
grammar corrections to the manuscript, and C. Guzm\'an for
computer assistance.



\begin{thebibliography}{99}

\bibitem{} Abel T., Bryan G. \& Norman M.L., 2000, ApJ, 540, 39

\bibitem{} Abgrall H., Roueff E., Viala Y., 1982, A\&ASS, 50, 505

\bibitem{} Barkana R., Loeb A., 2001. Phys.Rep., 349, 125

\bibitem{} Bromm V. \& Clarke C. J., 2002, ApJ, 566, L1

\bibitem{} Bromm V. \& Larson R. L., 2004, ARA\&A, 42, 79

\bibitem{} Bromm V., Coppi P.S. \& Larson R. B., 2002, ApJ, 564, 23  

\bibitem{} Ciardi B, \& Ferrara A., 2004, Space Science Reviews, in print

\bibitem{} Dalgarno A. \& McCray R.A., 1972, AR\&A, 10, 375

\bibitem{} Dalgarno A. \& Roberge W. G., 1979, ApJ, 233, L25

\bibitem{} Flower D. R., 2000, MNRAS, 318, 875

\bibitem{} Flower D.R. \& Pineau des For\^ets G., 2001, MNRAS, 323, 672 

\bibitem{} Flower D. R. \& Roueff E., 1999, MNRAS 309, 833

\bibitem{} Flower D.R., Le Bourlot J., Pineau des Forets G. \& Roueff E.,
2000, MNRAS, 314, 753

\bibitem{} Galli D. \&  Palla F., 1998, A\&A, 335, 403

\bibitem{} \_\_\_\_\_\_., 2002, Planetary \& Sp. Sci., 50, 1197

\bibitem{} Le Bourlot J., Pineau des For\^ets G. \& Flower D.R., 1999, MNRAS,
305, 802

\bibitem{} Le Petit F.,  Roueff E. \& Le Bourlot J., 2002, A\&A, 390, 369

\bibitem{} Lepp S. \& Shull J.M., 1984, ApJ, 280, 465

\bibitem{} Nakamura F. \& Umemura M., 2002, ApJ, 569, 549 

\bibitem{} Mac Low M. M. \& Shull J.M., 1986, ApJ, 302, 89  

\bibitem{} Palla F., Galli D. \& Silk J., 1995, ApJ, 451, 44

\bibitem{} Puy D., Alecian G., Le Bourlot J., Leorat J. \& 
Pineau Des Forets G., 1993, A\&A, 267, 337

\bibitem{} Roueff E., Flower D.R., 1999, MNRAS, 305, 353

\bibitem{} Roueff E., Zeippen C.J., 1999, A\&A, 343, 1005

\bibitem{} Shapiro, P.R. \& Kang, H. 1987, ApJ, 318, 65

\bibitem{} Shaw G., Ferland G. J., Abel N. P., Stancil  P. C.,
\& van Hoof  P. A. M., 2005, ApJ, in press (astro-ph/0501485) 

\bibitem{} Stancil P. C.,  Lepp S. \& Dalgarno A., 1998, ApJ, 509, 1

\bibitem{} Tegmark M., Silk J., Rees M.J., Blanchard A., Abel T. \& 
Palla F., 1997. ApJ, 474, 1 

\bibitem{} Tin\'e S., Lepp S. \& Dalgarno A., 1998, in
``Memorie della Societa Astronomia Italiana'', vol 69, p.345

\bibitem{} Uehara H. \& Inutsuka S. 2000, ApJ, 531, L91  

\bibitem{} Varshalovich D.A., Khersonskii V.K., 1976, Sov. Astron. Lett., 2,
(6) 227

\bibitem{} Weil M. L., Pudritz R. E., 2001, ApJ, 556, 164

\end{thebibliography}
\end{document}